\documentclass[showpacs,prl,superscriptaddress,twocolumn,floatfix]{revtex4}
\usepackage{graphicx,float,amsmath,amssymb}

\def\I{{\rm i}}                  
\def\D{{\rm d}}                  
\newcommand{\smean}[1]{\langle #1 \rangle} 
\newcommand{\EXP}[1]{{\mbox{\large e}}^{#1}}         

\newcommand{\Lphi}{L_{\varphi}}

\begin{document}

\title{Geometrical dependence of decoherence by electronic \\interactions in 
       a GaAs/GaAlAs square network}

\author{M. Ferrier}
\affiliation{Laboratoire de Physique des Solides, Univ. Paris-Sud, 
             CNRS, UMR 8502, 91405 Orsay, France.}

\author{A. C. H. Rowe}
\affiliation{Laboratoire de Physique de la Mati\`ere Condens\'ee, 
           \'Ecole Polytechnique, CNRS, UMR 7643, 91128 Palaiseau, France.}

\author{S. Gu\'eron}
\affiliation{Laboratoire de Physique des Solides, Univ. Paris-Sud, 
             CNRS, UMR 8502, 91405 Orsay, France.}

\author{H. Bouchiat}
\affiliation{Laboratoire de Physique des Solides, Univ. Paris-Sud, 
             CNRS, UMR 8502, 91405 Orsay, France.}

\author{C. Texier}
\affiliation{Laboratoire de Physique Th\'eorique et Mod\`eles Statistiques,
              Univ. Paris-Sud, CNRS, UMR 8626, 91405 Orsay, France.}
\affiliation{Laboratoire de Physique des Solides, Univ. Paris-Sud, 
             CNRS, UMR 8502, 91405 Orsay, France.}

\author{G. Montambaux}
\affiliation{Laboratoire de Physique des Solides, Univ. Paris-Sud, 
             CNRS, UMR 8502, 91405 Orsay, France.}

\date{April 9, 2008}

\begin{abstract}
  We investigate weak localization in metallic networks etched in a two
  dimensional electron gas
  between $25\:$mK and $750\:$mK when electron-electron (e-e) interaction is
  the dominant phase breaking mechanism. 
  We show that, at the highest temperatures, the contributions arising from
  trajectories that wind around the rings and trajectories that do not are
  governed by two different length scales. 
  This is achieved by analyzing separately the envelope and the
  oscillating part of the magnetoconductance. For $T\gtrsim0.3\:$K we find
  $\Lphi^\mathrm{env}\propto{T}^{-1/3}$ for the envelope, and
  $\Lphi^\mathrm{osc}\propto{T}^{-1/2}$ for the oscillations, in agreement
  with the prediction for a single ring \cite{LudMir04,TexMon05}.
  This is the first experimental confirmation of the geometry dependence of
  decoherence due to e-e interaction.
\end{abstract}

\pacs{73.23.-b~; 73.20.Fz}



\maketitle


In a conductor, quantum electronic interference is limited by phase breaking
mechanisms and can only occur below a characteristic scale $\Lphi$, the
phase coherence length. Understanding $\Lphi$ is a fundamental issue of
mesoscopic physics. For samples much longer than $\Lphi$, regions of typical
size $\Lphi$ behave independently causing disorder averaging so that
only contributions of reversed interfering electronic
trajectories survive. This gives rise to a small quantum
correction to the average conductance, called the weak localization (WL)
correction, suppressed by a magnetic field. Thus the magnetoconductance
(MC) 
is a powerful experimental tool to measure $\Lphi$, determine its temperature
($T$) dependence and identify the scattering mechanisms responsible for
decoherence.
For a quasi 1D diffusive wire (of width $W\ll\Lphi$), the situation is now
well understood theoretically~\cite{AltAroKhm82} and
experimentally~\cite{experimentsLphi}~: when dephasing is dominated by
electron-electron (e-e) interaction, $\Lphi$ is well described by the
Altshuler-Aronov-Khmelnitskii (AAK) theory yielding $\Lphi\propto{T}^{-1/3}$.
For a ring threaded by a flux $\phi$, the WL oscillates as a function of
$\phi$ with a period $\phi_0/2$ due to interference between trajectories
enlacing the ring, where $\phi_0=h/e$ is the flux quantum. These are the
Altshuler-Aronov-Spivak (AAS) oscillations~\cite{AroSha87}. It was recently
pointed out in \cite{LudMir04,TexMon05} that decoherence due to e-e
interaction is geometry dependent~: AAS oscillations involve a length scale
$\Lphi$ different from the one of the AAK result for the wire.

The difference can be qualitatively understood along the following lines. The
WL correction is related to the Cooperon $\mathcal{P}(t)$ which sums the
contributions of closed interfering reversed trajectories $\mathcal{C}_t$ for
a time $t$.
E-e interaction can be described as a fluctuating electromagnetic field which
randomizes the phase $\Phi[\mathcal{C}_t]$ accumulated along $\mathcal{C}_t$.
Thus the WL correction to the conductivity can be written as
\begin{equation}
  \label{ZeWL}
  \Delta\sigma = -\frac{2 e^2 D}{\pi} 
  \int_0^\infty \D t\, \mathcal{P}(t)\,
  \smean{\EXP{\I\Phi[\mathcal{C}_t]}}_{V,\mathcal{C}_t}
\end{equation}
where the average $\smean{\cdots}_{V,\mathcal{C}_t}$ is taken over the
fluctuations of the electric potential $V$ and the closed diffusive
trajectories $\mathcal{C}_t$. In order to get a qualitative picture, it is
sufficient to consider the fluctuations of the phase
$\smean{\Phi^2}_{V,\mathcal{C}_t}$, related to the fluctuations of $V$ via
$\frac{\D\smean{\Phi^2}_V}{\D{t}}=\int\D\tau\smean{V(\tau)V(0)}_V$. The power
spectrum of the potential is given by the Johnson-Nyquist theorem,
$\int_0^{t}{\smean{V(\tau)V(0)}_V\D\tau}=2e^2TR_{t}$, for the resistance
$R_{t}=x(t)/(W\sigma_0)$ of a wire of length $x(t)$, the typical distance
explored by interfering trajectories for a time scale $t$~; $\sigma_0$ is the
Drude conductivity and $W$ the section of the wires. Since the scaling of
$x(t)$ with $t$ depends on the diffusion and therefore on geometry, from
$\frac{\D\smean{\Phi^2}_{V,\mathcal{C}_t}}{\D{t}}\sim\frac{e^2T}{W\sigma_0}x(t)$
we see that the decoherence depends on geometrical properties. For an infinite
wire, $x(t)\sim\sqrt{Dt}$, so that the phase fluctuation varies as
$\smean{\Phi^2}_{V,\mathcal{C}_t}\sim(t/\tau_N)^{3/2}$ \cite{MonAkk05}, with
Nyquist time $\tau_N\propto{T}^{-2/3}$. This well-known result of AAK yields
$L_\varphi\sim{L_N}=\sqrt{D\tau_N}\propto{T}^{-1/3}$
\cite{AltAroKhm82,AkkMon07}. However for a finite wire or for a ring of
perimeter $L$, the length $x(t)$ cannot be greater than $L$ at large times,
leading to fluctuations $\smean{\Phi^2}_{V,\mathcal{C}_t}\sim{}t/\tau_c$ with
the new time scale $\tau_c\propto{T}^{-1}$,
yielding~$L_\varphi\sim{L_c}=\sqrt{D\tau_c}\propto{T}^{-1/2}$~\cite{LudMir04,TexMon05}.
 
In a ring, the WL involves two kinds of trajectories: those which do not
enclose the ring whose dephasing is expected to be driven by $\tau_N$
and those which enclose the ring and necessarily explore the whole system so
that their dephasing is driven by $\tau_c$. In the Fourier series of the
conductivity, the harmonic $n=0$ (smooth part) corresponds to trajectories
which do not enlace the ring and are only affected by the penetration of the
magnetic field in the wires. Therefore we expect the $T$ dependence of the
envelope to be characterized by $\Lphi^\mathrm{env}\propto{T}^{-1/3}$, as in
an infinite wire. On the other hand, for harmonics $n\neq0$, trajectories
causing the AAS oscillations encircle at least once the ring. As a consequence
the harmonics decay for $\Lphi\ll{L}$, defined as
$\EXP{-nL/\Lphi^\mathrm{osc}}$ \cite{footnote0}, is characterized by 
$L_{\varphi}^\mathrm{osc}\propto{T}^{-1/2}$~\cite{footnote1}.
Thus {\it the envelope and the oscillations of the MC are controlled by two
  different length scales}.

In this letter, we show how these two length scales can be extracted from the
MC of networks fabricated from a GaAs/AlGaAs 2D electron gas
(2DEG). 
$\Lphi$ is de\-ter\-mi\-ned independently from the harmonics content of the
AAS oscillations ($\Lphi^\mathrm{osc}$) and from the MC envelope
($\Lphi^\mathrm{env}$) between $25\:$mK and $750\:$mK. The main result is that
$\Lphi^\mathrm{osc}$ follows a $T^{-1/3}$ law up to 300 mK and a $T^{-1/2}$
law above, whereas $\Lphi^\mathrm{env}$ follows a $T^{-1/3}$ law up to
$750\:$mK. This is the first experimental evidence that in the high $T$
regime, $\Lphi^\mathrm{env}/L\lesssim0.3$, the phase coherence length follows
a different power law with $T$ for harmonics $0$ and~$1$.


\begin{figure}[!ht]
\includegraphics[scale=0.75]{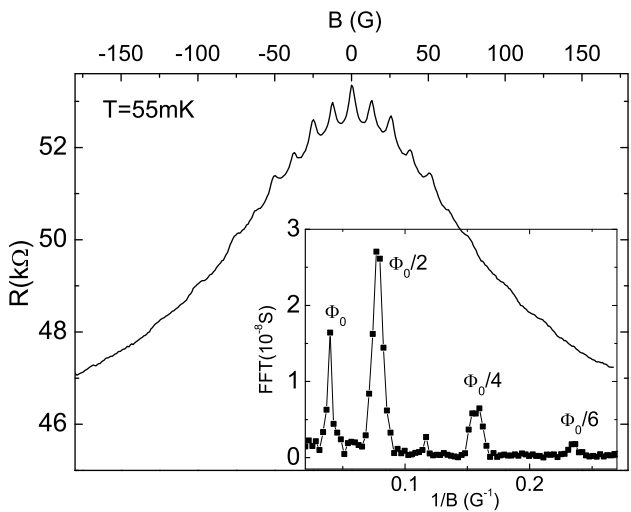}
\caption{ {\it Magnetoresistance of network C at $55\:$mK.} Inset~: {\it FFT
    of the MC after subtraction of the envelope. The first peak corresponds to
    Aharonov-Bohm oscillations.} }
\label{TF}
\end{figure}

Measurements were performed on two different networks described in
\cite{ferrier04}, and therein denoted sample~A (with electrostatic gate)
and sample~C (without). They consist of $10^6$ square loops of side
$a=1\:\mu$m for sample~A (a $1000\times1000$ grid connected at two opposite
corners through 100 wires) and $a=1.2\:\mu$m for sample~C (a rectangular grid
of aspect ratio 5 connected at the two narrow sides). 
The nominal width of the wires is $W_{0}=0.5\:\mu$m. We have measured the MC
up to $4.5\:$T between $25\:$mK and $1.3\:$K, using a standard lock-in
technique (current of $1\:$nA at $30\:$Hz). The samples were strongly depleted
at low $T$ because of the etching step used to define the network. The
intrinsic electron density of the 2DEG, $n_e=4.4\times10^{15}\:$m$^{-2}$ for
sample~A and $3.8\times10^{15}\:$m$^{-2}$ for C, is recovered by illuminating
the samples for several minutes at $4.2\:$K. The density was determined from
Shubnikov-de Haas oscillations visible above $1\:$T. The carrier mobility was
estimated to be $\mu=2.2\:$m$^{2}$V$^{-1}$s$^{-1}$, 10 times smaller than the
mobility of the original 2DEG.

A typical experimental curve is shown in Fig.~\ref{TF}. At low magnetic field,
the MC exhibits large AAS oscillations with a period corresponding to
$\phi_{0}/{2}$ per unit cell (allowing a precise determination of $a$). At
$55\:$mK, three harmonics are visible in the Fourier spectrum of the MC (inset
of Fig.~\ref{TF}). The oscillations are damped above $60\:$G, but the field
dependence of the envelope, due to the penetration of the magnetic field
through the wires, is still clearly visible. At high $T$, the AAS oscillations
gradually disappear even at low field and only the positive MC remains with a
smaller amplitude. The first harmonic is detectable up to $750\:$mK and the
second up to $350\:$mK.


In \cite{ferrier04} only the ratio between first and second harmonics was
exploited below $350\:$mK, whereas the present work reports new data in a 
broader temperature range extending into a regime where $\Lphi\ll{L}=4a$. 
Difficulties arise due to uncertainties in some sample parameters such as
the real width $W$ of the wires or the elastic mean free path $\ell_e$. 
This is why we have developed the following strategy 
to extract $\Lphi$ as a function of $T$ (and other parameters) with the least
possible parameters. At high $T$ ($\gtrsim1\:K$) decoherence is dominated by
electron-phonon interaction and is well described in eq.~(\ref{ZeWL}) by a
simple exponential
$\smean{\EXP{\I\Phi[\mathcal{C}_t]}}_{V,\mathcal{C}_t}=\EXP{-t/\tau_\varphi}$
independent on trajectories, where $\tau_\varphi=\Lphi^2/D$. In this case the
theoretical calculation for the WL correction to the conductance of a square
network~\cite{DouRam85,ferrier04} 
$
\Delta{G}=C\,\Delta\tilde\sigma(\Lphi,\phi)=
C\,[\Delta\tilde\sigma_\mathrm{env}(\Lphi)+\Delta\tilde\sigma_\mathrm{osc}(\Lphi,\phi)]$
perfectly describes the experimental
results~\cite{DouRam85,PanChaRamGan84,Mal06}. $\phi=Ba^2$ is the flux per cell
and $\Delta\tilde\sigma=W\Delta\sigma$ (so that $\Delta{G}$ and
$\Delta\tilde\sigma$ do not depend explicitely on $W$).
$\Delta\tilde\sigma_\mathrm{env}$ and $\Delta\tilde\sigma_\mathrm{osc}$
designate the smooth and oscillating (AAS) parts, respectively. $C$ depends on
the network and its connection to contacts (from classical combination of
resistances in series and in parallel we find that the classical Drude
conductance is $G_D\simeq\frac{\sigma_0W}{3.5\,a}$ for geometry~A, therefore
$C^\mathrm{A}\simeq1/(3.5\,a)$~; for geometry~C we have
$C^\mathrm{C}\simeq1/(5\,a)$).
At low $T$ ($\lesssim1\:K$), decoherence is dominated by e-e interaction and
depends on nature of trajectories. Therefore we analyze separately the
envelope and oscillating part of MC. The envelope calculated with exponential
relaxation interpolates between
$\Delta\tilde\sigma_\mathrm{env}\simeq-\frac{2e^2}{h}\Lphi$ for $\Lphi\ll{a}$
and $-\frac{2e^2}{h}\frac{a}\pi\ln(\Lphi/a)$ for $\Lphi\gg{a}$, and is not
expected to depend much on the precise modelization of decoherence. On the
other hand it was shown~\cite{ferrier04,Mal06} that the experimental result is
perfectly fitted by the MC curve $\Delta\tilde\sigma_\mathrm{osc}$ with
exponential relaxation. Therefore we analyze the experiment with~:
\begin{equation}
\label{dG}
\Delta G= C_0\,\Delta\tilde\sigma_\mathrm{env}(\Lphi^\mathrm{env})
         +C_1\,\Delta\tilde\sigma_\mathrm{osc}(\Lphi^\mathrm{osc},\phi)
\:.
\end{equation}
To compare our measurements with the theory (\ref{dG}) we need $5$ parameters:
$\Lphi^\mathrm{osc}$, $\Lphi^\mathrm{env}$, $W$, $C_0$ and $C_1$. To begin
with, we extract directly $\Lphi^\mathrm{osc}$ (without any other adjustable
parameter) below $400\:$mK using the ratio of the two first harmonics of the
MC~\cite{ferrier04}. Using this result we determine the prefactor $C_1$ and
eventually extract $\Lphi^\mathrm{osc}$ from the amplitude of the first
harmonic up to $750\:$mK. Then we can determine the width of the wires from
the adjustment of the damping of the oscillating part. Finally we obtain
$\Lphi^\mathrm{env}$ and $C_0$ for the whole range of $T$ by fitting the MC
envelope with only these two parameters.

\begin{figure}[!ht]
\includegraphics[scale=0.75]{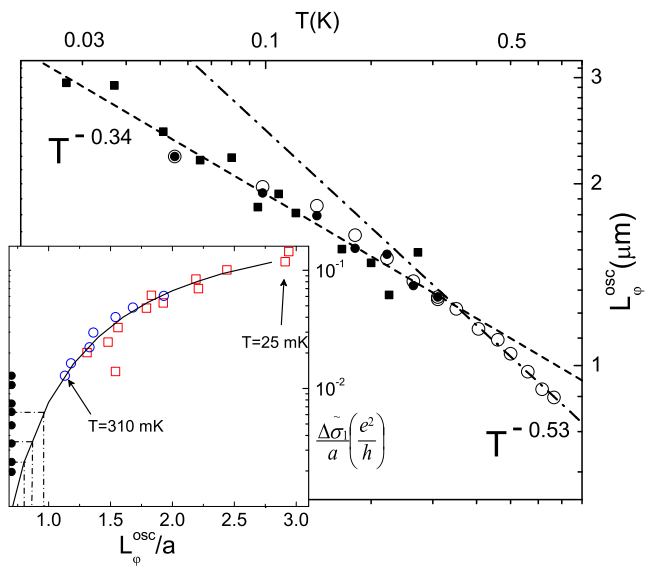}
\caption{{\it $\Lphi^\mathrm{osc}(T)$ for samples A (squares) and C (circles).
    Filled symbols are obtained from the ratio of the two first harmonics~;
    open symbols are obtained with only the first harmonic. The dashed line is
    a fit of the low $T$ part giving a $T^{-0.34}$ law and the dash-dotted
    line is the high $T$ fit yielding a $T^{-0.53}$ law.} Inset : {\it
    continuous line is the numerical calculation of $\Delta\tilde\sigma_1/a$
    as a function of $\Lphi/a$. Also plotted~: $\Delta G_1/(C_1a)$ for
    experimental data (open squares for A and open circles for C). Filled
    circles on the Y axis : experimental values of $\Delta G_1/(C_1a)$ for
    sample C measured at high $T$ when $\Lphi^\mathrm{osc}$ is unknown. As
    shown on the graph $\Lphi^\mathrm{osc}$ can be deduced from the
    corresponding abscissa on the theoretical curve.} }
\label{sigma}
\end{figure}

As explained in~\cite{ferrier04}, we extract $\Lphi^\mathrm{osc}$ by comparing
the experimental MC harmonics [$\Delta{G}_n(T)$] to the theoretical harmonics
$\Delta\tilde\sigma_n(\Lphi^\mathrm{osc})$. Experimental Fourier peaks are
integrated to account for the penetration of the magnetic field in the wires.
At low $T$ ($\lesssim300\:$mK), when the second harmonic
$\Delta{G}_2$ is clearly visible, this allows a direct determination of
$\Lphi^\mathrm{osc}$ since the harmonics  ratio
$R_{12}=\frac{\Delta\tilde\sigma_1}{\Delta\tilde\sigma_2}=\frac{\Delta{G}_1}{\Delta{G}_2}$
depends only on $\Lphi^\mathrm{osc}/a$.
%
It yields $\Lphi^\mathrm{osc}(T)\propto{T}^{-0.34\pm0.02}$ (Fig.~\ref{sigma}),
in agreement with~\cite{ferrier04}.

At higher temperature, $\Delta{G}_2$ is suppressed and the $T$ dependence of
$L_{\varphi}^\mathrm{osc}(T)$ can only be extracted from $\Delta{G}_1$.
This requires the knowledge of $C_1$, given by
$\Delta{G}_1(\Lphi^\mathrm{osc})=C_1\Delta\tilde\sigma_1(\Lphi^\mathrm{osc})$.
Below $300\:$mK, since $\Lphi^\mathrm{osc}(T)$ and $\Delta{G}_1(T)$ are
determined, we plot $\Delta{G}_1$ as a function of $\Lphi^\mathrm{osc}/a$ and
superimpose experimental data with theoretical calculation for
$\Delta\tilde\sigma_1(\Lphi^\mathrm{osc})$ between $25\:$mK and $300\:$mK
(inset of Fig.~\ref{sigma}). This yields $C^\mathrm{A}_1\simeq1/(25a)$ and
$C^\mathrm{C}_1\simeq1/(20a)$. This plot shows that $C_1$ does not depend on
$\Lphi$ and consequently on $T$, which validates our assumption.
It is now possible to compare directly the experimental value of
$\Delta{G}_1(T)$ with the theoretical
$C_1\Delta\tilde\sigma_1(\Lphi^\mathrm{osc})$ (inset of Fig.~\ref{sigma}) for
the whole $T$-range where oscillations are visible and reconstruct the curve
$\Lphi^\mathrm{osc}(T)$ up to $700\:$mK. The result (Fig.~\ref{sigma}) clearly
shows that at high temperature $\Lphi^\mathrm{osc}(T)$ is no longer described
by the same power law. The fit of the high $T$ data gives
$L_{\varphi}^\mathrm{osc}\propto{T}^{-0.53\pm0.05}$, very close to the
expected $T^{-1/2}$ for an individual ring~\cite{LudMir04,TexMon05}. It is
interesting that the crossover between the two different power
laws occurs for $\Lphi^\mathrm{osc}/a\simeq1.2$, precisely when the second
harmonic becomes unobservable. This is consistent with the fact that the
network result coincides with the single ring result only when the probability
for trajectories to wind around two cells is negligible~\cite{footnote1}.

\begin{figure}[!ht]
\includegraphics[scale=0.9]{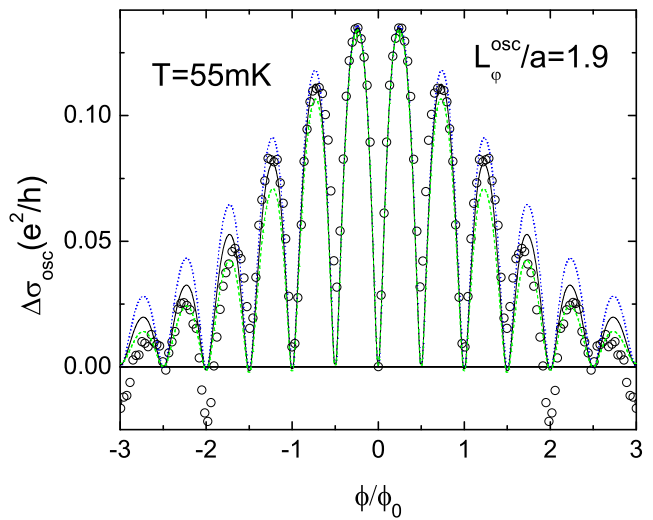}
\caption{\it Reconstruction of the damping of the oscillations after
  subtracting the envelope. Circles are experimental data for sample C at
  $55\:$mK. Lines correspond to theory for $W_\mathrm{eff}=0.07a$ (dotted),
  $0.08a$ (solid) and $0.09a$ (dashed). }
\label{FigW}
\end{figure}

Finally, the penetration of the magnetic field in the wires can be described
with the substitution $\Lphi\to\Lphi^\mathrm{eff}(\phi)$~\cite{AltAro81},
performed for both $\Lphi^\mathrm{env}$ and
$\Lphi^\mathrm{osc}$~\cite{footnote2}~:
\begin{equation}
  \label{lphidephi}
  \Lphi^\mathrm{eff}(\phi)=
  \Lphi/
  \sqrt{ 1 + \frac13
         \left(
            2\pi \frac{W_\mathrm{eff}\Lphi}{a^2} \frac{\phi}{\phi_0}
         \right)^2 
       }
  \:.
\end{equation}
$W_\mathrm{eff}=W\sqrt{\frac{3W}{9.5\ell_e}}$ (for $\ell_e\gg{W}$ and specular
reflections) is the width renormalized by the phenomenon of flux
cancellation~\cite{vantout88}. This penetration causes the damping
of the oscillations and the external MC envelope. First, we extract
$W_\mathrm{eff}$ from the damping of the oscillations. Since at large
$\phi$, $\Lphi^\mathrm{eff}$ is independent on $\Lphi$, this determination of
$W_\mathrm{eff}$ is very reliable. For a given $\Lphi^\mathrm{osc}$, we
compute the conductivity
$\Delta\tilde\sigma_\mathrm{osc}(\Lphi^\mathrm{osc}(\phi),\phi)$. Comparing
this curve to experimental data yields $W_\mathrm{eff}=80\pm5\:$nm
(Fig.~\ref{FigW}). For consistency, we have checked that
$W_\mathrm{eff}$ does not depend on $T$ from $25\:$mK to~$700\:$mK.

\begin{figure}[!ht]
\includegraphics[scale=0.95]{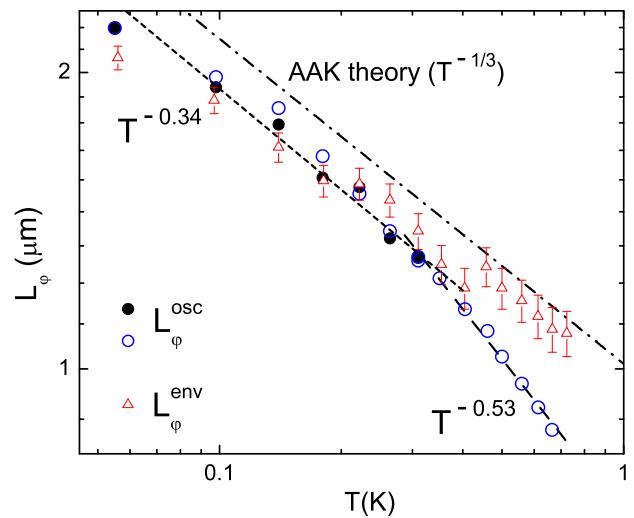}
\caption{\it $\Lphi$ vs $T$ for sample C. Symbols are those of
  Fig.~\ref{sigma}. Empty triangles are the values extracted from the MC
  envelope for sample C. Dashed lines are the fit of Fig.~\ref{sigma} giving
  the power laws $T^{-0.34}$ and $T^{-0.53}$. The dash-dotted line is the AAK
  prediction for a wire corresponding to our parameters. }
\label{Lphi}
\end{figure}

We then extract $\Lphi^\mathrm{env}$ from the MC envelope by a fitting
procedure. The theoretical expression of the envelope 
\cite{DouRam85,ferrier04} depends only on $W_\mathrm{eff}$,
$\Lphi^\mathrm{env}$ and $C_0$. Knowing $W_\mathrm{eff}$, only two
parameters remain, making the fit reliable. Fitting the experimental MC
envelope at each $T$ we get $C_0^\mathrm{A}\simeq1/(7\,a)$,
$C_0^\mathrm{C}\simeq1/(14\,a)$ and the curve $\Lphi^\mathrm{env}(T)$ plotted
in Fig.~\ref{Lphi}. The fact that $C_0$ does not depend on $T$ confirms the
reliability of our method. The curve $\Lphi^\mathrm{env}(T)$ yields a power
law $\Lphi^\mathrm{env}\propto{T}^{-0.34}$ for the whole $T$-range.

From $W_\mathrm{eff}$, $C_0$ and the measurement of
$G_D=C_0\frac{e^2}{h}k_FW\ell_e\simeq1/(46\:$k$\Omega)$ we obtain
$\ell_e\simeq310\:$nm and $W\simeq190\:$nm, which agrees with~\cite{ferrier04}.
The fact that we find $C_0^\mathrm{A}\simeq1/(7\,a)$ and
$C_0^\mathrm{C}\simeq1/(14\,a)$, instead of the expected
$C^\mathrm{A}\simeq1/(3.5\,a)$ and $C^\mathrm{C}\simeq1/(5\,a)$, shows that
there probably are a significant number of broken wires~\cite{ferrier04}.
This may partly explain the decoupling of the envelope
and the oscillating part of the MC ($C_0\neq{C}_1$). However we expect that
the broken wires have little effect on the determination of $\Lphi$ since the
non trivial shape of the MC is still very well described by the calculation of
$\Delta\tilde\sigma_\mathrm{osc}$ that does not include this
effect~\cite{footnote3}.
In the absence of a full theory for networks including  the effect of e-e
interaction, we based our analysis on the theoretical result (\ref{dG}) for
exponential relaxation of phase coherence instead of the single ring
result~\cite{LudMir04,TexMon05}. This emphasizes nonlocal properties of
quantum transport. However, we have not taken into account the fact that the
preexponential factor of harmonics is expected to be proportional to
$\Lphi^\mathrm{env}$ and not $\Lphi^\mathrm{osc}$~\cite{TexMon05}. This may
also partly explain the difference between $C_0$ and~$C_1$.

We now discuss the quantitative agreement of $\Lphi^\mathrm{env,osc}$ with
theories. AAK \cite{AltAroKhm82,AkkMon07} give
$\Lphi^\mathrm{AAK}=\sqrt{2}(\frac{D^2m^*W}{\pi{k_B}T})^{1/3}\simeq0.88a\,T^{-1/3}$,
with $T$ in Kelvin (Fig.~\ref{Lphi})~; $m^*$ is the effective mass.
$\Lphi^\mathrm{env}$ and $\Lphi^\mathrm{osc}$ are related by
$\Lphi^\mathrm{osc}/a=\varpi(\Lphi^\mathrm{env}/a)^{3/2}$~\cite{LudMir04,TexMon05,footnote0}
with $\varpi=\frac{2^{5/4}}{\pi}\simeq0.75$. We have obtained experimentally
$\Lphi^\mathrm{env}/a=0.81\,T^{-1/3}$ and
$\Lphi^\mathrm{osc}/a=0.62\,T^{-1/2}$ giving $\varpi\simeq0.85$ close to
theory.


In conclusion we have shown that decoherence due to e-e interaction is
geometry dependent. This has been revealed by demonstrating that, in the high
$T$ regime ($\Lphi^\mathrm{env}/a\lesssim1.2$), the envelope of the MC and its
oscillating part (AAS) are governed by different length scales
$\Lphi^\mathrm{env}\propto{T}^{-0.34\pm0.02}$ and
$\Lphi^\mathrm{osc}\propto{T}^{-0.53\pm0.05}$. These $T$-dependences are close
to the expected $\Lphi^\mathrm{env}\propto{T}^{-1/3}$~\cite{AltAroKhm82} and
$\Lphi^\mathrm{osc}\propto{T}^{-1/2}$~\cite{LudMir04,TexMon05}.
Prefactors found in experiment are also consistent with theories. We emphasize
that it has been possible to use the power law
$\Lphi^\mathrm{osc}\propto{T}^{-1/2}$, derived theoretically for a single
ring, in the regime $\Lphi^\mathrm{env}/a\ll1$ when rings of the network can
be considered as independent. In the other regime
$\Lphi^\mathrm{env}/a\gtrsim1.2$, the observed
$\Lphi^\mathrm{osc}\propto{T}^{-0.34}$ is still
unexplained~\cite{footnote1}.


We thank B.~Etienne and D.~Mailly for the fabrication of the heterojunctions
and L.~Angers, C.~B\"auerle, R.~Deblock, F.~Mallet, A.~Mirlin, L.~Saminadayar
and F.~Schopfer for fruitful discussions.



\begin{thebibliography}{100}
\bibitem{footnote1} Our analysis is only valid in the high $T$ regime
  ($\Lphi\ll{L}$). For $\Lphi\gtrsim{L}$, winding trajectories also explore
  the wires around the ring, which drastically affects the winding properties
  and leads to different behaviours 
  (for example, for a single ring connected
  to long leads one gets $\Lphi^\mathrm{osc}\propto{T}^{-1/6}$ for
  $\Lphi\gtrsim{L}$~\cite{TexMon05}).
\bibitem{footnote2} It was shown in~\cite{TexMon05} that the substitution for
  $\Lphi^\mathrm{osc}$ has another form for e-e interaction, however both
  functions are indistinguishable in practice.
\bibitem{footnote3}
  This is confirmed by the fact that a similar analysis holds for the silver
  networks of \cite{Mal06} with a negligible number of broken wires.
\end{thebibliography}
\end{document}